\documentclass{emulateapj}
\usepackage{graphicx}

\def\etal  {et~al.}
\shortauthors{Fixsen et al.} \shorttitle{The Extra-Galactic Sky
Temperature at 3-90 GHz}

\tighten

\begin{document}

\hfuzz=10pt
\overfullrule=0pt

\title{ARCADE 2 Measurement of the Extra-Galactic Sky Temperature at 3-90 GHz}

\author{ D.J. Fixsen\altaffilmark{1}, A. Kogut\altaffilmark{2},
S. Levin\altaffilmark{3}, M. Limon\altaffilmark{4}, P.
Lubin\altaffilmark{5},P. Mirel\altaffilmark{6}, M.
Seiffert\altaffilmark{3}, J. Singal\altaffilmark{7}, E.
Wollack\altaffilmark{2}, T. Villela\altaffilmark{8} and C. A.
Wuensche\altaffilmark{8}}

\altaffiltext{1}{University of Maryland,
                 Code 665, NASA/GSFC,
                 Greenbelt MD 20771.
         e-mail: dale.j.fixsen@nasa.gov}
\altaffiltext{2}{Code 665, Goddard Space Flight Center, Greenbelt,
MD 20771} \altaffiltext{3}{Jet Propulsion Laboratory, California
Institute of Technology, 4800 Oak Grove Drive, Pasadena, CA 91109}
\altaffiltext{4}{Columbia Astrophysics Laboratory, 550W 120th St.,
Mail Code 5247, New York, NY 10027-6902} \altaffiltext{5}{University
of California at Santa Barbara} \altaffiltext{6}{Wyle Information
Systems} \altaffiltext{7}{Kavli Institute for Particle Astrophysics
and Cosmology, SLAC National Accelerator Laboratory, Menlo Park, CA
94025} \altaffiltext{8}{Instituto Nacional de Pesquisas Espaciais,
Divis\~ao de Astrof\'{\i}sica, Caixa Postal 515, 12245-970 -  S\~ao
Jos\'e dos Campos, SP, Brazil} 

\begin{abstract}
The ARCADE~2 instrument has measured the absolute temperature of the
sky at frequencies 3, 8, 10, 30, and 90 GHz, using an open-aperture
cryogenic instrument observing at balloon altitudes with no emissive
windows between the beam-forming optics and the sky. An external
blackbody calibrator provides an {\it in situ} reference. Systematic
errors were greatly reduced by using differential radiometers and
cooling all critical components to physical temperatures
approximating the CMB temperature. A linear model is used to compare
the output of each radiometer to a set of thermometers on the
instrument. Small corrections are made for the residual emission
from the flight train, balloon, atmosphere, and foreground Galactic
emission. The ARCADE~2 data alone show an extragalactic rise of
$50\pm7$~mK at 3.3 GHz in addition to a CMB temperature of $2.730\pm
.004$~K. Combining the ARCADE~2 data with data from the literature
shows a background power law spectrum of $T=1.26\pm 0.09$~[K]
$(\nu/\nu_0)^{-2.60\pm 0.04}$ from 22~MHz to 10~GHz ($\nu_0=1$~GHz)
in addition to a CMB temperature of $2.725\pm .001$~K.
\end{abstract}

\section{INTRODUCTION}
The standard big bang model places the formation of the CMB at
$z\approx 6\times 10^6$ with a nearly perfect black body spectrum.
The black body spectrum remains in thermal equilibrium with the
electrons and ions in the early universe until the surface of last
scattering at z=1089. Measurements by the Far Infrared Absolute
Spectrophotometer (FIRAS) instrument across the peak of the CMB
spectrum ($\sim60$~GHz to $\sim600$~GHz) limit deviations from a
blackbody, with temperature 2.725$\pm$.001~K, to be less than 50
parts per million (Fixsen \& Mather 2002, Fixsen \etal\ 1996), but
measurements at centimeter or longer wavelengths are less
restrictive. Plausible energy-releasing processes including star
formation and particle decay or annihilation could produce
observable distortions at centimeter or longer wavelengths while
evading constraints at millimeter wavelengths.

At radio frequencies below 10~GHz, the radiation from the sky is
increasingly dominated by synchrotron and free-free emission both
from our own Galaxy and distant point sources and perhaps distant
diffuse sources. The ARCADE~2 (Absolute Radiometer for Cosmology,
Astrophysics and Diffuse Emission) instrument observes the CMB
spectrum at frequencies a decade below FIRAS to observe the spectrum
where the crossover occurs.

\section{THE INSTRUMENT}
The ARCADE~2 is a balloon-borne double nulled instrument with seven
radiometers at frequencies ranging from 3 to 90 GHz mounted in a
liquid helium bucket dewar. Each radiometer consists of cryogenic
and room temperature components. A cryogenic Dicke switch operating
at 75 Hz alternately connects the amplification chain to either a
corrugated horn antenna (Singal \etal\ 2005) or an internal
reference load (Wollack \etal\ 2007).  The temperature of the
reference load can be adjusted to produce zero differential signal,
nulling the radiometer output.  The horn in turn views either the
sky or an external blackbody calibrator.  The blackbody temperature
can be adjusted to match the sky temperature, nulling instrumental
offsets.

The calibrator (Fixsen \etal\ 2006) has reflection of less than -45
dB and can be positioned to fully cover the aperture of any of the
horns. Residual reflections from the calibrator are effectively
trapped within the horn/calibrator system. Within this system the
calibrator absorbs almost all of the radiation because the horn
emissivity is low and the calibrator emissivity is high. The effect
of reflection from the calibrator system is proportional to the
difference in temperature between the calibrator and the horn
antenna throat.

To minimize instrumental systematic effects the horns are cooled and
maintained at a nearly constant temperature ($\sim$1.5~K). The horns
have a $12^\circ$ full-width-half-maximum beam and are pointed
$30^\circ$ from the zenith to minimize acceptance of balloon and
flight train emission. A helium cooled flare reduces contamination
from ground emission. No windows are used. The ambient atmosphere
during flight is kept from the instrument by the efflux of helium
gas.

After the switch, a GaAs high electron mobility transistor (HEMT)
amplifier boosts the signal. The signal passes then through a
thermal break to a 280~K section where it is further amplified and
separated into two sub-bands followed by diode detectors, making
fourteen channels in all. Helium pumps and heaters allow thermal
control of the cryogenic components which are kept at 2-3.5~K during
the observations. The signal from each detector is demodulated with
a lockin amplifier operating synchronously with the Dicke switch.
The critical parameters of the radiometers and a full discussion of
the instrument are given in a companion paper (Singal \etal\ 2008).

The sky temperature measurement depends critically on the calibrator
temperature determination. The other components (horn, switch, cold
reference, and amplifier) become a transfer standard while comparing
the sky measurements to the calibrator measurements. There are 24
thermometers embedded in the calibrator, from the tips of the
calibrator cones to the back of the calibrator, along with 9 other
thermometers on other parts of the calibrator to measure the
temperature of its surroundings. These were included to look for
gradients and other artifacts as well as to provide redundancy in
the case of a thermometer failure. These ruthenium oxide
thermometers (Fixsen \etal\ 2002) are read out at 0.9375~Hz with
2~mK precision and 1~mK accuracy. Identical thermometers have been
calibrated on separate occasions over 5 years with absolute
calibrations stable to less than 2~mK.

\begin{figure*}[t]
\includegraphics[trim=0 40 0 0,width=7in]{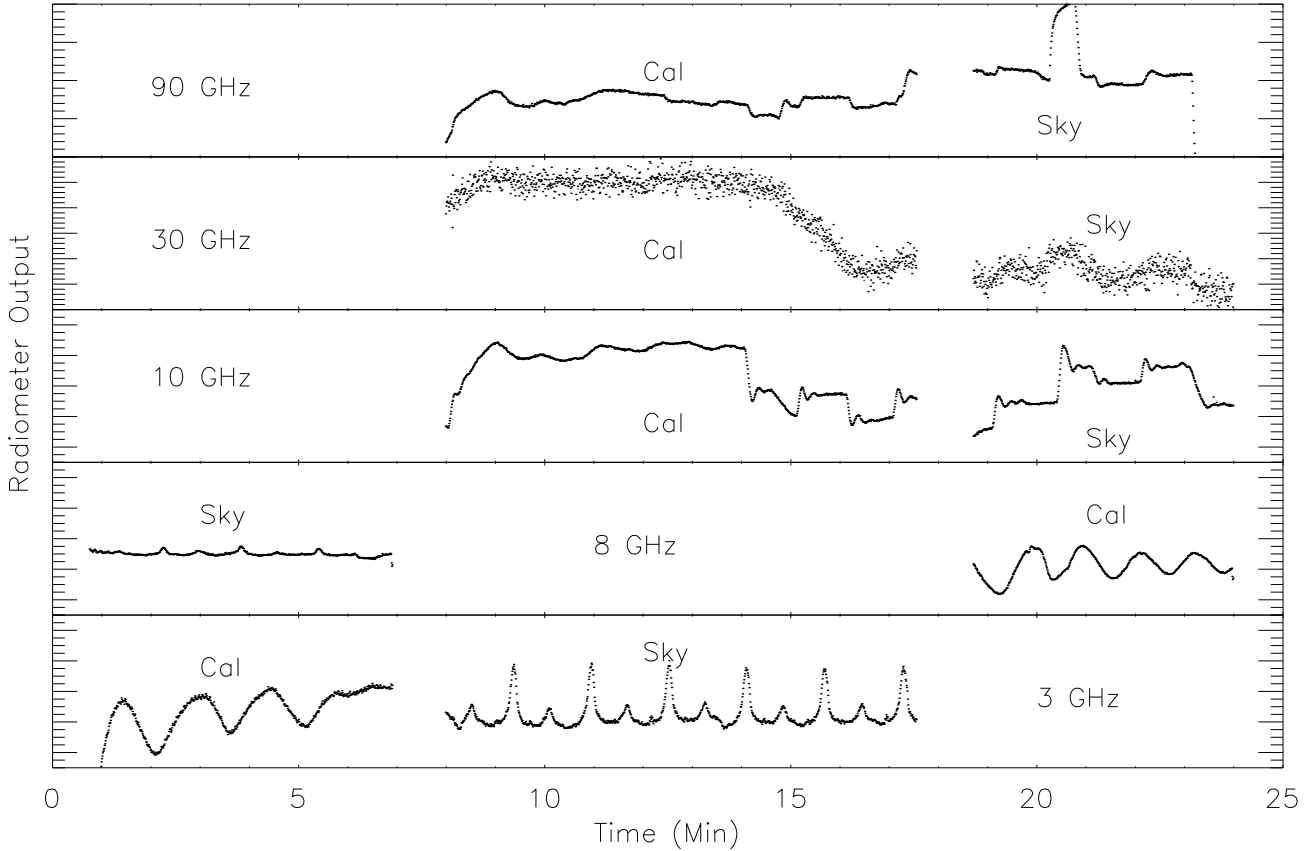}
\caption{A subset of raw data from the flight. Full scale on these
plots are $\sim$1~K for the 3,8, and 10~GHz radiometers, $\sim$3~K
for the 30~GHz radiometer and $\sim$2~K for the 90~GHz radiometer.
While the 3~GHz radiometer observes the calibrator, the 8~GHz
radiometer observes the sky etc. The 10, 30, and 90~GHz radiometers
view the sky or calibrator as a group. Intermediate transients have
been suppressed. The Galactic crossings are clearly evident in the
3~GHz and 8~GHz sky data. Reference load changes can be seen in both
the sky and calibrator in the 10~GHz and 90~GHz data.} \label{datas}
\end{figure*}

\section{THE OBSERVATIONS}
The ARCADE~2 instrument was launched from Palestine TX on a 29~MCF
balloon 2006 Jul 22 at 1:15~UT. The instrument reached a float
altitude of 37~km at 4:41~UT. The cover protecting the cryogenic
components was opened at 5:08 UT. The calibrator was moved 28 times
from 5:30 to 8:11 providing at least 8 cycles between calibrator and
sky for each of the radiometers. During this time the entire gondola
with the instrument was rotated at $\sim 0.6$~rpm, observing 8.4\%
of the entire sky.

The 5~GHz switch failed in flight, so there are no useful data from
that radiometer. The 30~GHz radiometer with the narrower beam is not
matched to the beams of the other radiometers and has much higher
noise than the other 30~GHz radiometer so its data are not used
here.

The most useful observations were from 5:35 to 7:40~UT and all of
the following derivations will use various subsets of this data.
During this time the calibrator temperature was controlled between
2.2 and 3.1~K with a mean temperature of 2.72~K. A selection of 25
minutes of raw data from the flight(approximately 10\% of the useful
data) are shown in Figure \ref{datas}. Only one of the channels for
each radiometer is shown. The other channel is similar. Full scale
on these plots are $\sim$1~K for the 3,8, and 10~GHz radiometers,
$\sim$3~K for the 30~GHz radiometer and $\sim$2~K for the 90~GHz
radiometer. While the 3~GHz radiometer observes the calibrator, the
8~GHz radiometer observes the sky and the 10, 30 and 90~GHz
radiometers ``observe" the flat aluminum underside of the carousel.
The 10, 30 and 90~GHz radiometers observe the sky and calibrator
together. The Galactic crossings are clearly evident in the 3~GHz
and 8~GHz sky data. Reference load changes can be seen in both the
sky and calibrator in the 10~GHz and 90~GHz data. The 30~GHz
radiometer data have much higher intrinsic noise. A first
approximation to the sky temperature can be obtained by selecting an
interesting sky datum on the figure and moving across to an
appropriate calibration datum then reading off the calibrator
temperature for that datum.

\section{SKY TEMPERATURE ESTIMATION}
Conceptually, the calibration process is straightforward. The
calibrator is placed over a radiometer horn, and the reference and
calibrator are each warmed and cooled to allow the measurement of
the emission coupling from each component into the radiometer. The
radiometer output is modeled as a linear combination of component
temperatures:
\begin{equation}
R\approx A\cdot T
\end{equation}
where $T$ a matrix of the relevant temperatures with each row a
component and each column a time. The radiometer outputs are a
vector, $R$. The solution, $A$, contains the couplings to the
various parts of the radiometer. Since the lockin has both a
positive and negative phase the couplings can be either positive or
negative. The coupling parameters include the gain and well as the
emissivity. The calibrator is then moved away so the radiometer
observes the sky; the parameters measured while observing the
calibrator can then be used to deduce the temperature of the sky. As
the mathematics are developed, it is important to remember that the
essential comparison is between the sky and the calibrator which
brackets the sky temperature while the rest of the radiometer is in
a similar state.

The most efficient use of the data uses all of the component
temperature variations to obtain the best coupling estimates.
However, some of the variations occur while the radiometer is
observing the sky, which because of the beam scanning the sky
includes the Galactic variation as well. The data could be binned by
sky pixel and a solution made for each point, but that would not
take advantage of the instrument variations happening while
observing other pixels. Instead, a Galaxy model developed in a
companion paper (Kogut \etal\ 2008) is subtracted at each pixel so
only the uniform background sky remains. Then a general least
squares fit is used to solve for the emissivities, gain and the sky
background temperature simultaneously, including all of the sky
observations.

The Galaxy model is derived iteratively. For the first iteration the
Galactic model is zero. The residual calibrated time series,
combined with pointing data, is then used to generate a map as in
Figure 1 of Kogut \etal\ (2008). The multi-frequency sky maps are
combined to form a model of Galactic emission, which is then used to
correct the time-ordered data for subsequent iterations of the sky
solution. The Galaxy model depends on the {\it spatial variation}
while the sky background temperature depends on the {\it zero level}
so the convergence is vary rapid. Since the model here implicitly
assumes a uniform sky no variations are injected into the Galactic
model from the fit. However, the background sky temperature depends
on the {\it absolute} level of the Galactic model.

\subsection{Galactic Foreground Subtraction}
 The Galactic foreground is complicated. A companion paper (Kogut \etal\
2008) describes the detailed model produced using the ARCADE~2 data
at 3, 8, and 10 GHz along with published results from sky surveys at
higher and lower frequencies.  The spatial structure of Galactic
radio emission is modeled as a linear combination of template maps
based on the 408 MHz survey (Haslam \etal\ 1981) and the full sky
C{\sc ii} map from Fixsen \etal\ (1996).  The offset of the template
model is then adjusted to match the total Galactic emission toward a
set of reference lines of sight.

The distinction between Galactic and extragalactic radiation varies
from author to author.  We define the total Galactic emission along
selected lines of sight using two independent techniques.  The first
method treats the Galaxy as a plane-parallel structure, and bins the
temperature of the ARCADE~2 sky maps and low-frequency radio surveys
by the $\csc|b|$ to determine the Galactic emission at the north or
south galactic poles.  A second technique uses atomic line emission
to trace Galactic structure.  We correlate the ARCADE~2 or radio
data against the map of C{\sc ii} emission to determine the ratio of
radio to line emission in the interstellar medium.  We then multiply
this ratio by the observed C{\sc ii} intensity toward selected lines
of sight (north or south galactic poles plus the coldest patch in
the northern sky) to estimate the total radio emission associated
with the Galaxy along each line of sight.  The two techniques agree
well along each independent line of sight.  We then compare the
template model to the estimate of total Galactic emission along each
line of sight to derive a single offset required to force the
template model to match the model of total Galactic emission.  The
three lines of sight provide consistent estimates for the offset in
the template model, with scatter 5~mK at 3~GHz and less than 1~mK at
8 or 10~GHz.

\subsection{Calibrator Thermometry}
Selecting which thermometer to use for the calibrator in the least
squares solution would be trivial if there were only one thermometer
or all of the thermometers read the same temperature. Using many
thermometers in the equation allows the radiometer data themselves
to select the best linear combination to describe the radiometer
data. However the best differentiation of the various thermal modes
of the calibrator is data from the times that the calibrator is
moving or rapidly changing temperature. Unfortunately these data can
not be used as they are taken during the calibrator movement or when
the thermal state of the calibrator is poorly determined.

One way out of this dilemma is to make a thermal model of the
calibrator using all of the data, then use that model during the
times when the thermal and radiometric state of the calibrator is
best understood to calibrate the radiometer. This has the advantage
that the data can be used for the calibrator model even when the
calibrator is being observed by a different radiometer.

A straight-forward way of generating a model of the calibrator is to
use principle component analysis (PCA) to extract the most essential
calibrator modes. The calibrator has 24 thermometers embedded in the
absorber to measure its temperature. A set of 21 of these
thermometers was carefully recalibrated after the flight. The data
from these thermometers is arranged as a matrix $T$ of 21 rows of
measurements each 6996 measurements long, corresponding to over 2
hours of observation. The eigenvector decomposition is
\begin{equation}
V\cdot D\cdot V^T=T\cdot T^T
\end{equation}
where $D$ is a diagonal matrix of 21 eigenvalues and $V$ is a set of
21 eigenvectors such that $V\cdot V^T=I$. This set of eigenvectors
describes and organizes the data into modes. The corresponding
eigenvalue calibrates the importance of each mode.

One of the critical differences between these modes is the response
time of each mode. The best data to distinguish the time constants
is the first few seconds after a thermal shock such as the movement
of the carousel. The first eigenmodes exhibit the crucial thermal
dynamics while the last eigenmodes are residual noise.

If the calibrator were perfectly isothermal the first (largest)
eigenvalue would contain all of the variance and the corresponding
eigenvector would equally weight each of the thermometers. In fact
the largest eigenvalue has 99.9\% of the variance and the weights
for the various thermometers only varies by a $\sim 5$\% in the
eigenvector.

\begin{figure}
\includegraphics[trim=100 0 0 30,width=3.5in]{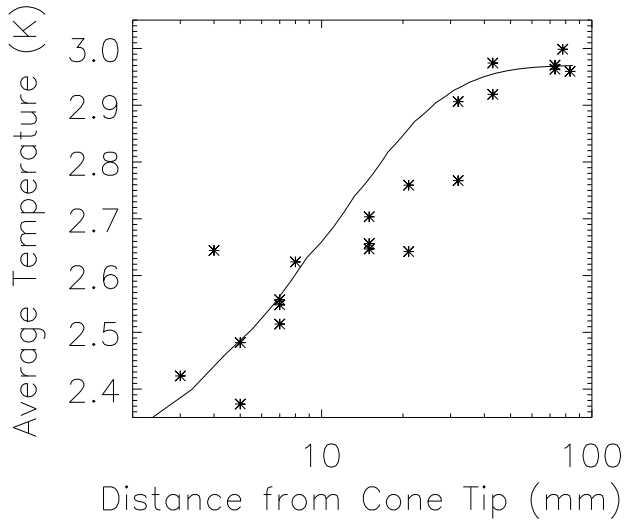}
\caption{The temperature within the calibrator averaged over the
data period verses the distance from the tip (point). The line is
the preflight prediction of the shape of the gradient. The preflight
prediction was used to select the placement of the thermometers. It
is not used in the analysis. The 21 thermometers are concentrated
near the tips to fully sample the gradient. Some of the dispersion
of the measured points is due to the radial gradient which is not
shown here.} \label{Cal}
\end{figure}

The second largest eigenmode (0.08\% of the temperature variance) is
a front to back gradient in the calibrator.  Such a gradient was
anticipated based on heat flow from the 2.7~K calibrator to the
colder (1.5~K) aperture below.  A wrap-around tank of superfluid
liquid helium surrounds the back and sides of the calibrator
structure to intercept heat from the outside.  A thermally
conductive aluminum shield lies between the tank and the absorbing
cones to provide an isothermal surface at approximately 2.7~K.  The
face of the calibrator opens to an aluminum plate maintained at the
bath temperature (1.5~K).  While the calibrator does not touch the
plate, the diffuse ($\sim~300$~Pa) helium gas column can transport
heat from the absorbing cones to the plate.  The resulting heat flow
creates a thermal gradient within the calibrator, which we measure
using thermometers embedded within the absorbing cones.

Figure \ref{Cal} compares the measured gradient to the locations of
the thermometers within the cones.  The total front to back gradient
is approximately 600~mK; however, since most of the gradient occurs
near the cone tips, 97\% of the absorber volume remains within 10~mK
of the base temperature.  The thermometer locations were chosen
using a simple static thermal model (Fixsen \etal\ 2006) and are
concentrated near the cone tips.  The thermometers are approximately
uniformly distributed along the actual gradient so that in-flight
gradients are well sampled throughout the absorber volume.

The details of the metal surface under the calibrator change as the
calibrator moves from one position to another.  We observe with the
calibrator in one of three positions.  While over the high frequency
horns, most of the calibrator is over a flat aluminum plate lying
1.5~mm below the cone tips.  A few individual cones lie over the
high-frequency horn antennas, with a larger gap between the cone tip
and the aluminum wall of the horn.  The 3~GHz horn is nearly the
same size as the calibrator.  While over the 3~GHz horn, most of the
cone tips are $\sim 150$~mm from the aluminum wall of the horn. The
third position has roughly half the cone tips near the aluminum
aperture plate and the other half suspended above the 5 and 8~GHz
horns.  The third and fourth eigenmodes show the imprint of the
large change in thermal conductivity between the cones and the 1.5~K
aperture corresponding to differences in the height of the gas
column below each cone. The thermal imprint of the three positions
can be considered three vectors.  Since the mean gradient has
already been removed, there remain only two dimensions.  The third
and fourth eigenmodes span this space. Together the first 4
eigenmodes account for 99.98\% of the temperature variance.

Smaller modes are more difficult to identify with known thermal
conditions but may reflect changes in helium flow or changes in the
aperture temperature during different times of the flight. We
include the next 6 modes to be conservative. This accounts for
99.996\% of the variation of the thermometers. The residual is
roughly consistent with noise. Thus we can describe the thermal
state of the calibrator with
\begin{equation}
U=V'\cdot T
\end{equation}
where $V'$ is $V$ truncated to 10 rows.

\subsection{The Solution}
A linear model can then be used to predict the radiometer output:
\begin{equation}
R=g E\cdot X
\end{equation}
where $g$ is the responsivity of the radiometer, $X$ is the matrix
of ten thermal modes, $U$,(each row is a mode, each column a time)
augmented by rows for the sky temperature, the reference load
temperature, the horn temperature, the switch temperature, and a
fifth order polynomial. $E$ is a vector of emissivities and $R$ is a
vector of radiometer readings. Since the radiometer is followed by a
lockin the sign of $E$ is positive for the horn and calibrator but
negative for the reference load. Some components (eg. the switch or
some of the gradients) can have either sign depending on the details
of the unwanted asymmetries. Since neither $g$ nor $E$ is known a
priori, they are combined; $A=gE$. A least squares fit:
\begin{equation}
A=(X\cdot W\cdot X^T)^{-1}\cdot X^T\cdot W\cdot R
\end{equation}
produces the weighted solution to the optimum parameterization $A$,
where $W$ is a weight matrix. The solution contains the sky
background temperature as well as the gain and other parameters of
the fit.

To minimize extrapolation and possible nonlinearities the calibrator
temperature range should match the sky temperature range. The
reference and horn temperatures should either match the sky
temperature or remain stable. Table \ref{meantemps} shows the mean
temperatures and their variations for the flight times when the data
are used. The calibrator temperature listed is the simple average of
the calibrator thermometers.

\begin{table}[tbph]
\begin{center}
\begin{tabular}{lcccc}
\multicolumn{5}{c}{Mean Temperatures and RMS
Variations}\\
 Radiometer & Calibrator & Horn Antenna & Reference &
HEMT Amp
\\\hline
 3 GHz Rad & 2731$\pm$134 & 1486$\pm$3 & 1987$\pm$48 & 1439$\pm$3 \\
 8 GHz Rad & 2710$\pm$116 & 1414$\pm$3 & 1474$\pm$3 & 1440$\pm$3 \\
10 GHz Rad & 2728$\pm$111 & 1470$\pm$3 & 2840$\pm$158 & 1403$\pm$3 \\
30 GHz Rad & 2728$\pm$111 & 1635$\pm$379 & 2290$\pm$737 & 1436$\pm$3 \\
90 GHz Rad & 2724$\pm$108 & 2775$\pm$173 & 2970$\pm$349 & 2961$\pm$784 \\
\hline
\end{tabular}
\caption{\label{meantemps} Mean temperatures and RMS variations of
the major components of the radiometers. Different radiometers
observed the calibrator at different times. All temperatures in the
table are in milliKelvin.}
\end{center}
\end{table}

The average calibrator temperature is well matched to the
temperature of the sky, and the variation in temperature of the
calibrator covers the same range as the variation in sky
temperature, thus the estimation of the sky temperature is an
interpolation rather than an extrapolation. This also means that the
thermodynamic temperature scale of the calibrator is carried to the
sky and no further corrections are needed to make the result a
thermodynamic temperature.

Equation (5) assumes a linear model of radiometer coupling. Such an
assumption is well justified.  The power at the detector diode for
each radiometer is dominated by the noise temperature of the cold
amplifier. Even for the radiometer with the lowest noise temperature
(8~K), the largest Galactic signal variations of 0.15~K represent
less than 2\% of the total detector loading.  Furthermore, the mean
difference in power between the calibrator observations and the sky
observations is less than 1\%. Gain compression and non-linear
effects are negligible for such small variations.

As can be seen from Table~\ref{meantemps}, the mean temperatures of
the major components are near the CMB temperature. This minimizes
the effects of reflections, unmodeled emission and responsivity
variations. The cold reference has as much impact on the radiometer
signal as the sky or calibrator. But the sky temperature estimation
does not depend on the {\it absolute} accuracy of the reference
thermometer. The reference load and the rest of the instrument are
merely a transfer standard to compare the calibrator to the sky.
Nevertheless the reference thermometer and all of the other
thermometers are read out to a precision of 2~mK and have been
calibrated to 2~mK against an absolute NIST standard.

While the horns are considerably cooler than the sky, their
temperatures are very stable.  Hence whatever signals they
contribute during sky measurements are repeated during the
calibration, and as they are included in the least squares fit, no
further correction is needed. The temperature of these elements is
set by the vapor pressure of superfluid helium. Since the horns are
coated with a film of superfluid helium they are isothermal. The
small variations in temperature are driven by the changes in balloon
altitude. The reference load temperatures for 3 and 8 GHz are low
compared to the sky and calibrator, leading to a large signal in the
radiometer output. This can allow instrument gain variation to
affect the inferred sky signal. Gain variations are measured through
the temperature variation of the calibrator. They are removed with a
fifth order polynomial.

All of the data are weighted equally and assumed to be independent,
except some data are excised by making the weight, $W$, zero. This
excising is done to eliminate the data that are obviously bad or
suspect on grounds other than their position within the residual
distribution. For example, during part of the flight the lower 3~GHz
band shows a signal at one azimuth. This signal is not seen in any
of the other channels. This is precisely the type of signal one
would expect from a radar watching the balloon with a narrow
frequency beam. These data were excised. An additional 9\% of the
data were excised as outliers; most of these measurements were taken
near the edge of the Galaxy where there is a high spatial gradient
and pointing errors as well as the details of the beam shape are
most important. These data have a minimal effect on the background
estimation.

In addition to the thermometers in the calibrator, references, and
horns there were thermometers on the HEMT amplifiers. Since the HEMT
amplifiers follow the Dicke switches, they cannot affect the offset
of the radiometers, but their gain can affect the output. To correct
for any temperature dependence in the gain, the row for the
amplifier in the temperature matrix contains $R*\delta T_{amp}$
where $\delta T_{amp}=T_{amp}-\left< T_{amp}. \right>$ The mean of
the temperature is removed to improve the condition of the matrix
which would otherwise have a row nearly identical to the data being
fit.

\begin{figure*}[t]
\includegraphics[trim=0 40 0 0,width=7in]{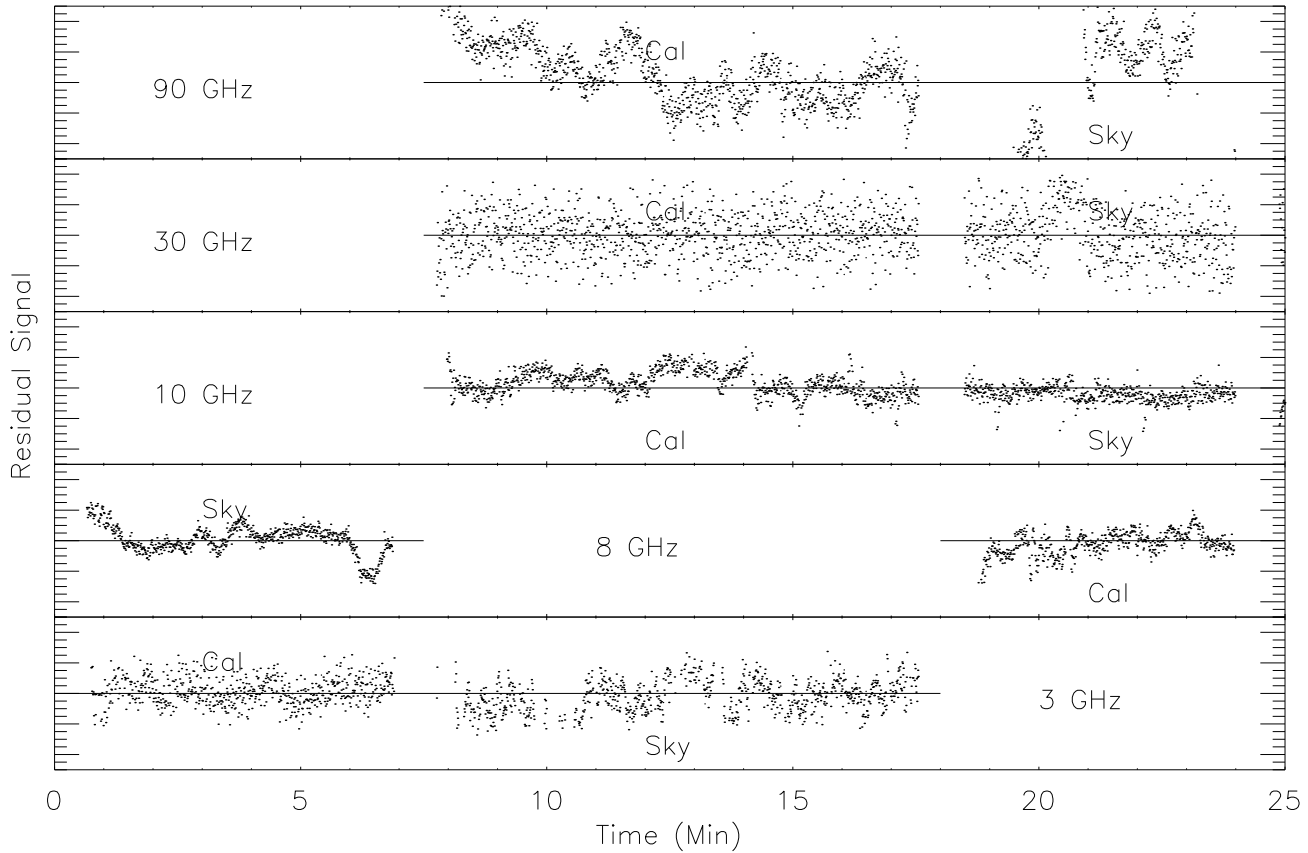}
\caption{The residuals of the fit to the data from figure
\ref{datas}. Full scale is +/-~50~mK for all radiometers except the
30~GHz which is +/-~500~mK. Excised data are not shown. Sky and
calibrator observations are labeled.} \label{residuals}
\end{figure*}

Figure~\ref{residuals} shows the residuals in the data after the
model has been removed. Comparing Fig.~\ref{datas} with
Fig.~\ref{residuals} it can be seen that while the radiometer
component temperatures vary by $\sim 150$~mK, the residuals of the
fit vary by only tens of milliKelvins. This demonstrates that the
system is close to linear, and that all of the major components are
accounted for.

\subsection{Instrumental Foreground Estimation}
Most of the instrument was in the far sidelobes of the antenna beams
so its thermal emission to the radiometer is negligible. However the
flight train, consisting of the parachute, ladder, FAA transmitter
and balloon is directly above the instrument 30$^\circ$ from the
center of the beam. Its emission could not be ignored. Since the
flight train is complicated and moves with the balloon rather than
the gondola, a reflector constructed of aluminum foil covered foam
board was attached to the gondola to hide these components and
instead reflect the sky into the radiometers. The signal from these
local sources is calculated by Singal \etal\ (2008).

One of the principal advantages of a balloon flight is that it puts
the instrument above about 99.7\% of the atmosphere and an even
larger fraction of the water vapor. The residual atmosphere
contributes less than 1~mK to the 10~GHz channel (Staggs, 1996) and
a smaller amount for lower frequencies. The atmospheric signal is
too small to be seen in our tipping scans; a correction is made and
a 30\% uncertainty is included in the final uncertainty estimate.

\begin{table}[tbph]
\begin{center}
\begin{tabular}{lrrrrrrrrrr}
\tablewidth{7.5in} Component&3L&3H&8L&8H&10L&10H&30L&30H&90L&90H \\
\hline
Instrument&10.8&5.8&36.6&42.2&2.9&2.3&4.4&4.8&17.2&16.9\\
Atmosphere&0.7&0.7&0.7&0.7&0.7&0.7&2.2&2.2&5.8&5.8\\
Galaxy&23.2&19.0&2.2&1.9&1.2&1.1&.08&.07&0&0\\
\hline
\end{tabular}
\caption{\label{fore} Estimates of foreground radiation.  All
estimates are in mK from Singal \etal\ (2008).}
\end{center}
\end{table}

A second advantage of a balloon flight is it gets the instrument to
35 km, well above the nearest source of any radio transmitters.
Although the instrument was sensitive to nearby radio transmitters
the sensitivity was substantially reduced when the transmitter was
below the plane of aperture. While the radio noise in a city might
be significant the balloon trajectory explicitly avoids cities and
spends only a few minutes over small towns. Other than the radar
signature in one channel we see no evidence for any ground based
radio interference, so other than throwing out the apparently
contaminated data we make no correction for radio interference.

The horns are corrugated to limit the electric field near the wall.
The gap between the calibrator and the aperture plate is only 2\% of
a wavelength at 3~GHz. The radiation leaking through this gap was
measured to be $\sim$-50~dB. The induced signal on the radiometer is
less than .1 mK. At higher frequencies the gap is a larger faction
of a wavelength but the edge is much further away so the net effect
is a much smaller leakage at higher frequencies.

\section{UNCERTAINTY ESTIMATION}
Operating the radiometer in a near null condition allows precise
measurements to be made with greatly relaxed constraints on the
gain, linearity and reflection of the system. It is instructive to
imagine an ideal situation where all of the components of the
radiometer (horn, calibrator, switch, reference, amplifier, etc) are
at the same temperature as the sky. In this case there is no change
in radiometer output when switching from sky to calibrator, and the
gain, offset and linearity of the radiometer are irrelevant.
Instrumental reflections do not matter since signals reflected into
the radiometer have the same temperature as the sky. What matters in
this ideal case is only the calibrator temperature and the
contributions of foregrounds.

ARCADE~2 was operated within 0.1 K of ideal, except for the horns
and the references of the 3 and 8~GHz radiometers. The horns have a
mean temperature 1.22 K below the CMB temperature, but the horn
temperature is very stable. The reference temperatures of the 3 and
8~GHz are too low, requiring a polynomial fit to allow for gain
variation during the observations. The range of calibrator
temperatures measured throughout the observations includes
significant overlap with the 2.72 K of the CMB.

The overall uncertainty in the radiometric temperature is a
combination of the uncertainties of the parts of the model that go
into the radiometric temperature estimate. Each of the uncertainties
listed in Table~\ref{uncertain} will be discussed in turn.

\begin{table}[tbph]
\begin{center}
\begin{tabular}{lrrrrrrrrrr}
\tablewidth{7.5in} Source&3L&3H&8L&8H&10L&10H&30L&30H&90L&90H\\
\hline
ThermometerCal&1.0&1.0&1.0&1.0&1.0&1.0&1.0&1.0&1.0&1.0\\
Radiometer Cal&6.7&5.7&4.2&4.4&4.3&4.2&153&75&35&20.0\\
Statistics&5.0&4.7&7.7&8.6&3.9&4.1&27.3&13.5&13.8&6.9\\
Galaxy&5.3&4.9&0.6&0.6&0.4&0.3&0.0&0.0&0.0&0.0\\
Inst~Emiss&3.2&1.7&11.0&12.7&0.9&0.7&1.3&1.4&5.2&5.1\\
Atmosphere&0.2&0.2&0.2&0.2&0.2&0.2&0.7&0.7&1.4&1.4\\\hline
Total&10.5&9.1&14.1&16.0&6.0&6.0&155&76.2&38&21.8\\
\hline
\end{tabular}
\caption{\label{uncertain}Uncertainty estimates are discussed in \S
5. Uncertainties are added in quadrature. All estimates are in mK.}
\end{center}
\end{table}

\subsection{Absolute Thermometer Calibration Uncertainty}
The sky temperature can not be determined to better accuracy than
the absolute calibration of the thermometers in the calibrator. The
thermometer calibration was tested several times before the flight.
After the flight 21 thermometers from the calibrator, still embedded
in  their cones, were carefully compared to a NIST calibrated
thermometer over the 2.2 to 3.6 K range (Singal \etal\ 2008). The
flight electronics and flight cables were used in the test and the
test was repeated 3 times to verify the uncertainties. In addition,
the lambda transition to superfluid helium at 2.17~K was clearly
seen with the NIST standard thermometer in the calibration data
providing an absolute reference. An estimate of the uncertainty for
this test is 1 mK. By using all 21 of the thermometers, the
individual errors are further suppressed, except for the uncertainty
of the NIST calibrated thermometer, which is common to all the
thermometers.

\subsection{Temperature Gradient Uncertainty}
The uncertainty in the sky temperature is dominated by thermal
gradients in the calibrator. If the calibrator were isothermal, its
only contribution to the sky temperature uncertainty would be the
absolute calibration uncertainty of the embedded thermometers.
Spatial gradients are observed within the absorber cones. The
largest gradient averages 600~mK front-to-back, with the absorber
tips cooler than the back. Transverse gradients are much smaller,
with a mean gradient of 20~mK. These gradients are not stable in
time, but vary slowly with scatter comparable to the mean amplitude.

The radiometric temperature of the calibrator depends on the
integral of the temperature distribution within the absorber,
weighted by the electric field at the antenna aperture. This
integral is approximated as a linear combination of ten thermal
modes of the calibrator. The time variation in the temperatures and
radiometer output is used to derive a single time-averaged weight
for each of the eigenmodes. The procedure is insensitive to thermal
gradients in directions not sampled by the thermometers, or on
spatial scales smaller than the spacing between thermometers.

Small scale gradients are not likely to be significant for three
reasons. First, heating is spread over the entire back of the
calibrator, and cooling is done by diffuse gas over the entire front
of the calibrator so the thermal dynamics admit only large scale
gradients. Second, the natural frequency of gradients is
proportional to the scale size to the inverse second power. Hence
any high spatial frequency gradients will be quickly damped. We
conservatively excise all data taken within 20 seconds of a
calibrator move. Third, the low frequency channels observe a large
section of the calibrator (the 3 GHz radiometer observes the entire
calibrator) so any residual gradients will be smoothed in the low
frequency channels.

Although the gradients are well measured the coupling to the various
modes for each radiometer must be determined from the data. The
radiometer noise enters into the determination of the coupling to
the various modes. Hence the uncertainty is large for the
radiometers with high radiometer noise (eg the 30 GHz radiometer).

Estimating the uncertainty due to the gradients in the calibrator
presents a challenge. Since the gradients are not static the
relative weights of the various modes in the fit will compensate for
the gradients. To estimate the uncertainty in this compensation, the
sky temperature solution is repeated replacing the ten most
significant calibrator thermal modes with the temperatures of ten
randomly selected thermometers. Solutions using 1000 different
random thermometer selections are used to generate 1000 sky
temperatures. The standard deviation of the distribution of derived
sky temperatures is related to the uncertainty of the sky
temperature. Since there are 21 well calibrated thermometers and
only ten are used to generate the trial estimates, the dispersion in
the sky temperature estimates is an overestimate of the final
uncertainty. On the other hand, the normal estimate of the
uncertainty of the mean of the 1000 selections is an underestimate
of the uncertainty as the samples are not independent. We
conservatively estimate the uncertainty due to thermal gradients is
half of the dispersion or 16 times the standard uncertainty of the
mean.

\subsection{Statistical Uncertainty}
The statistical uncertainty is derived directly from the data. After
the residuals are computed the $\chi^2$ is renormalized so that the
$\chi^2$/DOF is one. The ideal would have the same amount of data
for each radiometer, but the vagaries of balloon flight and random
noise leave the remaining degrees of freedom with a range of 4312 to
8169 with an average of 6932.

The in-flight noise is roughly in agreement with the preflight data
for the 3~GHz and 10~GHz radiometers. The in-flight noise is an
order of magnitude higher than the preflight measurements for the
30~GHz data. In laboratory tests we noted that there were many
settings for which the HEMT's would oscillate. We had attempted to
set the control voltages of the HEMT amplifiers in valleys of
stability. Apparently conditions drifted in flight leading to
oscillations and excessive noise in the 30~GHz radiometer. The noise
in the sky data and calibrator data are identical within their
uncertainties. The 8~GHz radiometer in-flight noise is higher than
the preflight noise by a factor of 4. But the in-flight noise is
dominated by low frequency noise due to the gain variations on the
large signal due to the low reference load temperature. The
in-flight noise of the 90~GHz radiometer is also high due to drifts
in the warm mixer and local oscillator. The measured in-flight
variance is statistically propagated to the sky temperature
uncertainty in table~\ref{uncertain}.

\subsection{Galactic Emission Uncertainty}
The extragalactic sky temperature is defined as the residual
remaining after subtracting a model of Galactic emission.  We define
the absolute temperature of Galactic emission along 3 independent
lines of sight using the mean of the temperatures derived from the
plane-parallel spatial morphology of the Galaxy or the observed
correlation between radio and atomic line emission.  The three
reference lines of sight provide consistent estimates for the total
Galactic emission, with scatter 5~mK at 3~GHz, 0.5~mK at 8~GHz, and
0.4~mK at 10~GHz.

\subsection{Instrument and Atmosphere Emission Uncertainty}
The emission from the reflector and the flight train was modeled and
measured and the two agree within the measurement uncertainties. The
careful measurement of the beam from the mouth of the antenna allows
a very complete model. The tipping tests demonstrate that the model
is essentially correct. The major uncertainty in the model is the
emissivity of the aluminum foil on the foam. By calibrating against
the measurement this uncertainty is reduced.

The minimal contribution of the atmosphere at 37~km is from Danese
\& Partridge (1989) and Liebe (1981). We assume a 30\% uncertainty
in both of these sources.

\subsection{Instrument Drifts}
Of some concern are the possible drifts of the instrument gain and
offset in the instrument. The offset of the high frequency
amplifiers is effectively canceled by chopping between the reference
and the sky/calibrator at 75~Hz. The gain of these amplifiers might
be temperature dependent. The cold HEMT amplifier is expressly
checked in the model, although excluding the amplifier temperature
from the fit does not result in a significant change in the final
temperature estimation. The warm amplifiers were cooling very slowly
during the observations. Including a linear gain drift did not
significantly improve the fit or alter the final temperature, and
thus it was not included in the final fit. The helium liquid level
changed over the course of the observations. But all of the
components of the radiometers were well above the liquid helium for
the entire set of observations.

A camera showing the exposed parts of the ARCADE instrument at
$\sim$2.5~K shows some nitrogen ice buildup on the insulators and
outer parts of the ARCADE~2 instrument. Ice here has no radiometric
effects. The radiometric contribution from nitrogen ice collecting
on the aperture plane and flares visible in our camera is negligible
as it remains at the temperature of the aperture plate or flare.

The efflux of 5 m$^3$ s$^{-1}$ of boiloff helium gas prevented
nitrogen ice from accumulating on the optics.  No condensation is
visible on the horn antennas.  If small amounts were to collect
within the horns, it would freeze out at the horn mouth since the
flow of helium gas from the horn impedes nitrogen flow into the
interior. Any nitrogen freezing on the horn will have negligible
radiometric impact since solid nitrogen has no rotational modes and
hence no emission lines at centimeter wavelengths.  Furthermore,
since it is in thermal contact with the horn it must remain close to
the horn temperature.  Its effects are limited to changing the
dielectric surface of the horn near its mouth, where the electric
field is largely decoupled from the walls.  Finally, whatever
minimal effect nitrogen snow might have will cancel in the
sky/calibrator comparison.

It is difficult for nitrogen to collect on the inside of the
calibrator as the back of the calibrator is sealed and the front of
the calibrator is closed by the aperture plane or the horn. Both of
these are well below the freezing point of nitrogen so any nitrogen
getting to the surface of the horn or aperture plane will freeze
immediately and stay where it first comes into contact with the
aperture plane or flare. The calibrator looks down so no nitrogen
snow or oxygen rain can fall into it.

Many voltage, current, and other temperatures (both cryogenic and
ambient) were tested for correlation with residuals. Such a test can
show some connection even if the connection is not immediately
understood. None of the auxiliary sensors showed any significant
correlations with the residuals.

\section{Extra-Galactic Spectrum}
\begin{figure}
\includegraphics[trim=100 0 0 30,width=3.5in]{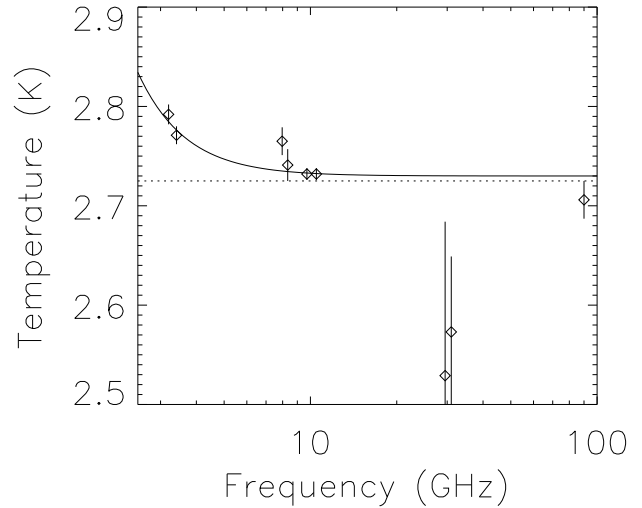}
\caption{The thermodynamic temperature as a function of frequency.
The solid line is the best fit to the ARCADE~2 data with a constant
CMB temperature plus a synchrotron like component with an assumed
-2.62 index. The vertical lines are $\pm 1 \sigma$. The dotted line
is the FIRAS CMB temperature.} \label{Arc}
\end{figure}

Figure \ref{Arc} shows the extragalactic spectrum measured by the
ARCADE~2 instrument.  Although the data at 10, 30, and 90~GHz are
consistent with the CMB temperature $2.725 \pm 0.001$~K measured by
the COBE/FIRAS instrument at frequencies above 60~GHz (Fixsen \&
Mather, 2002), the data at 8 and 3~GHz show a clear excess.  The
excess is statistically significant, with both 3~GHz channels lying
more than 5 standard deviations above the FIRAS value.

The ARCADE~2 data alone can not constrain the spectral dependence of
the excess signal to extrapolate to other frequencies. Additional
data from the literature were selected to compare to the ARCADE~2
data. Although there are many published measurements at frequencies
below 3 GHz, only a few have small enough beam and sufficient sky
coverage to separate the Galactic component from the extragalactic
component. We use surveys at 22 MHZ (Roger \etal\ 1999), 45 MHz
(Maeda \etal\ 1999), 408 MHz (Haslam \etal\ 1981), and 1420 MHz
(Reich \& Reich 1986) to estimate the Galactic and extragalactic
temperature.  As with the ARCADE~2 data, the total Galactic emission
is estimated along three reference lines of sight (north or south
Galactic poles plus the coldest patch in the northern Galactic
hemisphere) using both a csc$|b|$ model of the plane-parallel
spatial structure or the measured correlation between radio emission
from each survey and atomic line emission traced by the C{\sc ii}
survey.  The two methods agree well for the total Galactic emission
along each independent line of sight. The residual remaining after
subtracting the model Galactic emission from the measured radio
emission along each line of sight constitutes an extragalactic
background.  The scatter in this estimate from the three independent
lines of sight provides an estimate of the uncertainty in the
background temperature. Uncertainties in the gain and offset for
each survey contribute additional sources of uncertainty, which are
combined in quadrature.  Table \ref{data} summarizes the
extragalactic temperature derived from the ARCADE~2 data and
lower-frequency radio surveys. The largest uncertainty in the low
frequency data is the gain uncertainty, but the uncertainty in table
\ref{data} also includes the offset uncertainty and the uncertainty
in the Galaxy subtraction added in quadrature.

\begin{table}[tbph]
\begin{center}
\begin{tabular}{lccc}
\tablewidth{7.5in} &Frequency&Temperature&Uncertainty\\
Source&GHz&K&K \\ \hline
Roger&0.022&21200&5125\\
Maeda&0.045&4355&520\\
Haslam&0.408&16.24&3.4\\
Reich&1.42&3.213&.53\\ \hline
ARCADE~2&3.20&2.792&0.010\\
ARCADE~2&3.41&2.771&0.009\\
ARCADE~2&7.97&2.765&0.014\\
ARCADE~2&8.33&2.741&0.016\\
ARCADE~2&9.72&2.732&0.006\\
ARCADE~2&10.49&2.732&0.006\\
ARCADE~2&29.5&2.529&0.155\\
ARCADE~2&31&2.573&0.076\\
ARCADE~2&90&2.706&0.019\\
\hline
\end{tabular}
\caption{\label{data} Data used in the determination of the CMB and
low frequency rise estimates. The ARCADE~2 final measurements are
listed here along with their uncertainties. The low frequency
measurements are antenna temperature while the ARCADE~2 results are
thermodynamic temperature.}
\end{center}
\end{table}

\begin{figure}
\includegraphics[trim=100 0 0 30,width=3.5in]{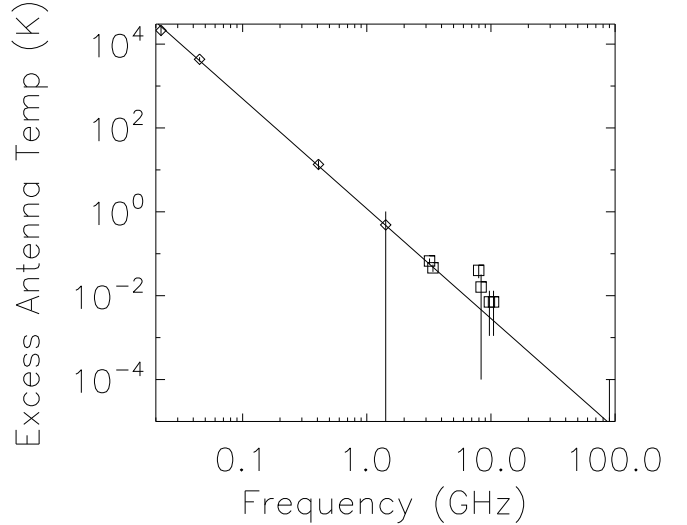}
\caption{The excess antenna temperature as a function of frequency.
The line is the best fit line with a -2.62 index. Diamonds are low
frequency points from the literature. Squares are ARCADE~2 data. The
30~GHz data point is included in the fit but since its excess
temperature comes out negative it does not appear on the plot. The
90~GHz error bar just appears at the lower right corner of the
plot.} \label{Sky}
\end{figure}

Inclusion of low-frequency radio surveys allows unambiguous
characterization of the excess signal in the ARCADE~2 data. The data
from Table \ref{Arc} are fit to the form
\begin{equation}
T(\nu) = T_0 + T_R ( \nu / \nu_0 )^\beta
\end{equation}
where $T_0$ is the CMB thermodynamic temperature and $T_R$ is the
normalization for a radio background. The radio background is
expressed in units of antenna temperature, related to the
thermodynamic temperature $T$ by
\begin{equation}
        T_A = T x /(e^x-1),
\end{equation}
where $x = h \nu / kT$, $h$ is Planck's constant, and $k$ is
Boltzmann's constant.  We obtain best-fit values $T_0 = 2.729 \pm
0.004$~K, $T_R  = 1.19 \pm 0.14$~K and $\beta = -2.62 \pm 0.04$ for
reference frequency $\nu_0 = 1$~GHz with $\chi^2=14.5$ for 10 DOF.
Figure \ref{Sky} shows the radio background after subtracting off
the best-fit CMB temperature. The ARCADE~2 data are in excellent
agreement with the radio background derived from the low-frequency
surveys.

\section{DISCUSSION}
\begin{table}[tbph]
\begin{center}
\begin{tabular}{lcccc}
\tablewidth{7.5in}Data
Sets&$T_R$(K)&Index&$T_0$(K)&$\chi^2$/DOF\\\hline
LF+ARC+FR&$1.26\pm 0.09$&$-2.60\pm 0.04$&$2.725\pm 0.001$&15.5/11\\
LF+ARC      &$1.26\pm 0.09$&$-2.62\pm 0.04$&$2.729\pm 0.004$&14.5/10\\
LF+FR       &$1.44\pm 0.41$&$-2.56\pm 0.10$&$2.725\pm 0.001$&1.0/2\\
ARC+FR         &$1.24\pm 0.15$&$-2.60$        &$2.725\pm 0.001$&14.2/8\\
LF             &$1.48\pm 0.53$&$-2.55\pm 0.10$&$2.6  \pm 0.6$&1.0/1 \\
ARC               &$1.13\pm 0.19$&$-2.60$        &$2.730\pm 0.004$&13.0/7 \\
\hline
\end{tabular}
\caption{\label{Solut} Various combinations of low frequency data
(LF), ARCADE~2 data (ARC), and FIRAS data (FR) are used to determine
the radio background and the temperature of the CMB. The FIRAS data
is treated as a single point with an effective frequency of
250~GHz.}
\end{center}
\end{table}
The ARCADE~2 measurement of the CMB temperature is in excellent
agreement with the FIRAS measurement at higher frequencies.  The
double-nulled design and novel open-aperture cryogenic optics
demonstrate significant improvements in both  calibration accuracy
and control of systematic errors compared to previous measurements
at these frequencies. With only two hours of  balloon flight
observations, ARCADE~2 approaches the absolute accuracy of
long-duration space missions.

The absolute temperature scale for ARCADE~2 is set by the
calibration of thermometers embedded in the external blackbody
calibrator, and is cross-checked using observations of the
superfluid transition in liquid helium. The largest uncertainties in
the ARCADE~2 measurements result from thermal gradients within the
blackbody calibrator. These gradients are driven by heat flow from
the 2.7 K calibrator through the diffuse helium gas to the colder
(1.5 K) aperture plate below. The temperatures within the calibrator
are monitored using 21 thermometers suitably spaced to uniformly
sample the calibrator gradient. The gradient is largely confined to
the tips of the absorber cones within the calibrator: 97\% of the
calibrator volume lies within 10~mK of the base temperature. A
principal component analysis of the thermometer data demonstrates
that the thermal state of the calibrator at any point in time can be
characterized using only a few modes formed from linear combinations
of the thermometers. The first 4 modes are clearly related to the
expected heat flow from the calibrator to the aperture, and account
for 99.98\% of the thermometer variance.  We conservatively model
the calibrator thermal state using the first 10 modes, accounting
for 99.996\% of the thermal variance. Tests comparing the sky
temperature derived after dropping individual thermometers
demonstrate that the calibrator has more than enough thermometers to
adequately sense the in-flight thermal gradients.  In fact, 4 or 5
well placed thermometers would have been sufficient to measure the
key thermal gradients within the calibrator.

Further improvements in the calibrator performance are possible. For
operational simplicity, the instrument design forces the temperature
of the aperture plate and horn antennas to the helium bath
temperature by flooding reservoirs attached to these structures with
superfluid liquid helium.  It would be straightforward to modify the
thermal design so that the aperture plate and horns are only weakly
coupled to the bath, allowing thermal control of these surfaces
analogous to the successful calibrator design.  Even modest thermal
control could reduce the temperature difference between the
calibrator and the aperture, thereby reducing heat flow and
associated thermal gradients within the calibrator by an order of
magnitude or more.

The detected extragalactic radio background is brighter than
expected. Low frequency Galactic radiation, and by extension
extragalactic radiation, is thought to be a mixture of synchrotron
and free-free emission.  Our analysis shows the detected background
to be consistent with a single power-law with spectral index $\beta
= -2.60 \pm 0.04$  from 22 MHz to 10 GHz.  Estimates of radio point
sources (Windhorst et al 1993, Gervasi et al 2008) indicate a
similar spectrum, but the radio background from ARCADE~2 and radio
surveys is a factor of $\sim 5$ brighter than the estimated
contribution of radio point sources.

It is difficult to reconcile the detected background with the
contribution from a population of radio point sources (Seiffert
\etal\ 2008).  Could the detected signal be in error?  The thermal
gradient in the ARCADE~2 calibrator is an obvious source of concern
for systematic errors.  However, the gradient is well sampled and
the uncertainties associated with the calibrator thermal state are
included in the ARCADE~2 uncertainties.  Furthermore, the bulk of
the gradient is concentrated at the tips of the cones.  The skin
depth for absorption within the calibrator is a function of
frequency: the high-frequency channels preferentially sample the
tips and outer surface of the absorber cones, while the 3 GHz
channel samples the entire absorber volume.  The results agree
within uncertainties with previous measurements at 10~GHz and 30~GHz
(Staggs \etal\ 1996, Fixsen \etal\ 2004). The agreement between the
high-frequency channels and the FIRAS CMB temperature argues against
any undetected systematic errors associated with thermal gradients
within the calibrator.

Further evidence against an error in the ARCADE~2 data comes from
comparing the ARCADE~2 data to a similar analysis using independent
radio surveys (Table \ref{data}).  Repeating the CMB/radio fit from
Eqn. 6 but using various combinations of data yields parameters in
table \ref{Solut}. None of the combinations are in serious
disagreement with any of the others. The FIRAS data constrain the
CMB temperature. The low frequency data constrain the index. The
ARCADE~2 data best constrain the radio background amplitude. The
ARCADE~2 data alone do not have sufficient low-frequency coverage to
determine the spectral index of the 3~GHz excess.  We assume a
spectral index -2.6 and fit the ARCADE~2 data alone or the ARCADE~2
and FIRAS data. The two independent data sets agree on both the CMB
temperature and radio amplitude, reducing the likelihood of serious
systematic error in either data set.

The extragalactic radio background is 2--3 times  as bright as the
Galactic emission toward the north or south polar caps.  Has
foreground emission been misinterpreted as background radiation?
Relativistic electrons trapped in the Earth's magnetic field emit
synchrotron radiation (Dyce \& Nakada 1959).  This emission is
polarized and anisotropic, however, with intensity peaking near the
Earth's magnetic equator and falling to zero at the poles.  Further,
it is measured to be less than 3~K at 30~MHz and expected to be much
less than 1~mK at 3~GHz (Ochs \etal\ 1963, Peterson \& Hower 1963).
It therefore is unlikely to be a significant contaminant of our
results. A spherical halo around the Galaxy with diameter comparable
to the disk would not show  up in the plane-parallel csc$|b|$
analysis, and would be identified with an extragalactic background.
However, the C{\sc ii} correlation analysis would identify any halo
containing ionized carbon.  The agreement of the csc$|b|$ and C{\sc
ii} techniques would indicate that any halo contribution is faint
compared to emission associated with the dominant plane-parallel
structure.  A similar radio/atomic line correlation using a
simultaneous fit to the C{\sc ii}, H$\alpha$, and H{\sc i} lines
shows no significant shift in the results, nor does infrared dust
emission show evidence for a significant halo.  A radio-bright halo
without dust, hydrogen, or carbon would be very peculiar.  Radio
observations of edge-on spiral galaxies typically show modest
high-latitude structure more typical of radio spurs extending from
the disk than a true spherical halo (Irwin et al. 2000, Irwin \etal\
1999, Hummel \etal\ 1991).  H$\alpha$ observations of edge-on
spirals indicate that diffuse high-latitude gas contributes only
12\% of the total H$\alpha$ intensity (Miller and Veilleux 2003),
well short of the factor of  2--3 required to explain the results of
this paper.  Finally, we note that estimates of the
Galactic/extragalactic separation along three independent lines of
sight agree on the background amplitude within 10\%, despite
Galactic foregrounds that vary by more than a factor of 2 from one
line of sight to another.

\acknowledgements This research is based upon work supported by the
National Aeronautics and Space Administration through the Science
Mission Directorate under the Astronomy and Physics Research and
Analysis suborbital program.The research described in this paper was
carried out in part at the Jet Propulsion Laboratory, California
Institute of Technology, under contract with the National
Aeronautics and Space Administration. We also wish to thank the
interns who worked on this experiment Adam Bushmaker, Jane Cornett,
Sarah Fixsen, Luke Lowe, and Alexander Rischard. T.V. acknowledges
support from CNPq grants 466184/00-0, 305219-2004-9 and
303637/2007-2-FA, and the technical support from Luiz Reitano.
C.A.W. acknowledges support from CNPq grant 307433/2004-8-FA.

\end{document}